\documentclass[10pt]{article}
\usepackage[OE]{express}
\begin{document}
\title{Stabilization of self-mode-locked quantum dash lasers by symmetric dual-loop optical feedback}
\author{Haroon Asghar,\authormark{*} Wei Wei, Pramod Kumar, Ehsan Sooudi, and John. G. McInerney\authormark{}}
\address{\authormark{}Department of Physics and Tyndall National Institute, University College Cork, Ireland T12 YN60}
\email{\authormark{*}haroon.asghar@ucc.ie} 
\begin{abstract}
We report experimental studies of the influence of symmetric dual-loop optical feedback on the RF linewidth and timing jitter of self-mode-locked two-section quantum dash lasers emitting at 1550 nm. Various feedback schemes were investigated and optimum levels determined for narrowest RF linewidth and low timing jitter, for single-loop and symmetric dual-loop feedback. Two symmetric dual-loop configurations, with balanced and unbalanced feedback ratios, were studied. We demonstrate that unbalanced symmetric dual loop feedback, with the inner cavity resonant and fine delay tuning of the outer loop, gives narrowest RF linewidth and reduced timing jitter over a wide range of delay, unlike single and balanced symmetric dual-loop configurations. This configuration with feedback lengths 80 and 140 m narrows the RF linewidth by $\sim$ 4-67x and $\sim$ 10-100x, respectively, across the widest delay range, compared to free-running. For symmetric dual-loop feedback, the influence of different power split ratios through the feedback loops was determined. Our results show that symmetric dual-loop feedback is markedly more effective than single-loop feedback in reducing RF linewidth and timing jitter, and is much less sensitive to delay phase, making this technique ideal for applications where robustness and alignment tolerance are essential.
\end{abstract}
\ocis{(140.4050) Mode-locked lasers; (140.3425) Laser stabilization; (270.2500) Fluctuations, relaxations, and noise.} 


\section{Introductions}
{Quantum nanostructure based mode-locked semiconductor lasers are of increasing interest for many applications as frequency comb sources in data centers [1], optical clock recovery [2] and high capacity coherent terabit communication systems [3]. While picosecond pulses from these lasers have been demonstrated routinely, these pulses have significant chirp and poor timing jitter. The latter is usually determined by measuring the linewidth of the repetition-rate peak in the RF intensity fluctuation spectrum. Several techniques such as single-loop external optical feedback [4-8], coupled optoelectronic oscillators (OEOs) [9-11], hybrid mode-locking [12], injection-locking [13-15] and dual-loop feedback [16-20] have been proposed and demonstrated to improve this key parameter of mode-locked lasers (MLLs). Among feedback-based techniques, optoelectronic feedback requires optical-to-electrical conversion, while hybrid mode-locking requires high speed electrical modulation of the gain and/or absorber sections. On the other hand, optical injection based techniques require a stable external laser, making these techniques less attractive in practice where low cost, simplicity and reliability are paramount. Of all stabilization techniques, external optical feedback is the simplest and most cost-effective demonstrated to date both experimentally [4-8] and numerically [21-24]. Most recently, 99\% reduction in RF linewidth and 23 fs pulse-to-pulse jitter was reported using single cavity feedback for a 40 GHz quantum-dot MLL [8]. Five different feedback regimes were identified along with the regime of resonant optical feedback most favorable and desirable for practical applications. In this demonstration [8], linewidth and timing jitter were shown to be very sensitive to small delay adjustments, with optimum performance being limited to one narrow regime (<15 ps delay). In practice, MLLs require reduced sensitivity of RF linewidth and timing jitter to detuning and drift in the delay phase. In particular, changing delay length should not cause switching into unstable or unwanted dynamical regimes.}\par
{Recently we have achieved lowest RF linewidth and reduced timing jitter over the widest delay range, introducing balanced asymmetric dual-loop (unequal arms of external loops) feedback such that the delay time of the exterior (shorter) loop was set to half that of the interior (longer) loop [20]. However, for ease of production and reliability of stable robust MLLs, using dual-loops with no or minimal delay difference is more advantageous and desirable. In this paper we propose symmetric dual-loop (SDL) feedback (equal arms of external loops), and demonstrate its efficacy over a much wider feedback range, for two-section self-mode-locked quantum dash (QDash) lasers emitting at $\sim$ 1550 nm and operating at $\sim$ 21 GHz pulse repetition rate. The feedback ratio for narrowest RF linewidth and lowest timing jitter was determined for single and SDL feedback schemes. We demonstrate that unbalanced SDL with fine tuning of the delay of the weaker feedback cavity (lowest feedback strength) produces narrowest RF spectra and reduced timing jitter across the widest delay range, unlike single-loop and balanced SDL feedback. Under stable resonant conditions and feedback level (-22 dB), RF linewidth is reduced from 100 kHz free-running to 3 kHz for single-loop feedback, 1.5 kHz for unbalanced SDL with 80 m loop and 1 kHz (instrument limited) for 140 m loop length. Moreover, RMS timing jitter is also reduced from 3.9 ps free-running to 0.6 ps for single-loop, 0.45 ps for unbalanced SDL with 80 m loop, 0.4 ps for 140 m loop (integrated 10 kHz-100 MHz). Our proposed unbalanced SDL scheme provides an effective regime of resonant feedback parameters much wider than single-loop and balanced SDL feedback, making it ideal for practical applications.}
\section{Experimental Setup}
{Devices under investigation were two-section InAs/InP QDash MLLs with active regions consisting of nine InAs quantum dash monolayers grown by gas source molecular beam epitaxy (GSMBE) embedded within two barrier layers (dash-in-barrier device), and separate confinement heterostructure layers of InGaAsP, emitting at $\sim$ 1.55 $\mu$m [25]. Cavity length was 2030 $\mu$m, 11.8\% (240 $\mu$m) of which formed the absorber section, giving pulsed repetition frequency $\sim$ 20.7 GHz (\(I_{Gain}\) = 300 mA, \(V_{Abs}\): Floating). Gain and absorber sections were electrically isolated by 9 k$\Omega$. The lasers were mounted p-side up (substrate down) on an AlN submount and a copper block with active temperature control. Electrical contacts were formed by wire bonding, and heat sink temperature was fixed at \(19^0\)C. Mode-locking was obtained without reverse bias applied to the absorber section. This was a two-section device but was packaged similarly to a single-section self-mode-locked laser since the absorber was unbiased: its minimal absorption does not affect the self-mode-locking mechanism [13]. Recently, 1.55 $\mu$m InAs/InP based QDash single-section self-mode-locked lasers have demonstrated promising high speed, narrow pulse generation, specifically GHz pulse repetition rate and very low RF linewidth [26, 27] with a low timing jitter. The absence of any obvious active/passive mode-locking scheme in these devices looks surprising at first glance. However, strong four-wave mixing [25] in the cavity has been proposed as the reason for this coherent self-pulsing behavior.}\par
     \begin{figure}[ht!]
\centering\includegraphics[width=11cm]{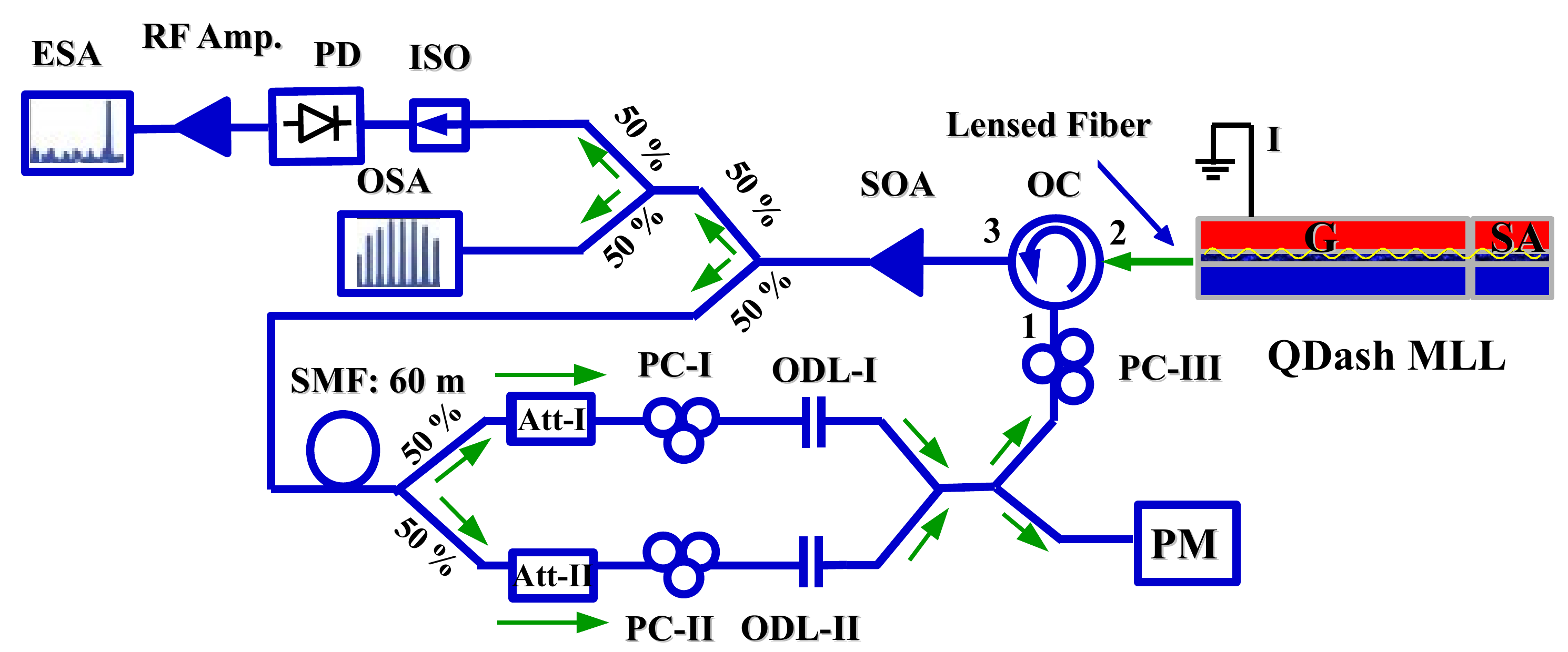}
\caption{Schematic of the experimental arrangement for single and dual-loop configurations. \textit{Acronyms}-- OC: Optical circulator; SOA: Semiconductor optical amplifier; ISO: Optical isolator; PD: Photodiode; ODL: Optical delay line; Att: Variable optical attenuator; PC: Polarization controller; SMF: Single mode fibre; ESA: Electric spectrum analyzer; OSA: Optical spectrum analyzer; PM: Power Meter; QDash MLL: Quantum dash mode-locked laser; G: Gain Section; SA: Saturable absorber}
\end{figure}
{Our experiment is depicted in Fig. 1. For single and dual-loop feedback, a calibrated fraction of light was fed back through port 1 of an optical circulator, then injected into the laser cavity via port 2. Optical coupling loss between adjacent ports was -0.64 dB. The output of the circulator was sent to a semiconductor optical amplifier (SOA, gain 9.8 dB) then split into two arms by a 50/50 coupler. Half the amplified signal went to an RF spectrum analyzer (Keysight E-series, E4407B) via a 21 GHz photodiode, and to optical spectrum analyzers (Ando AQ6317B and Advantest Q8384). The other half of the power was directed to the feedback arrangement. For a single feedback loop, all power passed through loop-I in Fig. 1. For SDL configurations, power was split equally into two loops via a 3-dB splitter, each containing an optical delay line, a variable optical attenuator and polarization controller. For SDL configurations, two combinations of feedback ratios were studied. For SDL with balanced feedback ratios, equal power was coupled to both external cavities. To implement unbalanced SDL feedback, more power (-20 dB) was coupled to loop-I than to loop-II (-26 dB). The lengths of the fiber loops were fine-tuned by optical delay lines based on stepper-controlled stages with delay resolution 1.67 ps. Polarization controllers in each loop and one polarization controller before port 1 of the circulator ensured the light from both loops matched the emitted light polarizations to maximize feedback effectiveness.}\par
 {RMS timing jitter is calculated from the single sideband (SSB) phase noise spectra measured for the fundamental RF frequency ($\approx$ 20.7 GHz) using:
  \begin{equation}
\sigma_{RMS}=\frac{1}{2 \pi f_{ML}}\sqrt{2 \int_{f_{d}}^{f_{u}} L(f)\,df}
\end{equation}
where \(f_{ML}\) is the pulse repetition rate, \(f_{u}\) and \(f_{d}\) are the upper and lower integration limits. \textit{L(f)} is the single sideband (SSB) phase noise spectrum, normalized to the carrier power per Hz. To measure RMS timing jitter of the laser in more detail, single-sideband (SSB) noise spectra for the fundamental harmonic repetition frequency were measured. To assess this, RF spectra at several spans around the repetition frequency were measured from small (finest) to large (coarse) resolution bandwidths. The corresponding ranges for frequency offsets were then extracted from each spectrum and superimposed to obtain SSB spectra normalized for power and per unit of frequency bandwidth. The higher frequency bound was set to 100 MHz (instrument limited).}\par
{We investigated the effects of three key parameters: external feedback level, optical loop length (80 and 140 m) and optical delay phase tuning, on the timing stability of our QDash MLL. The laser was subjected to single-loop and SDL (for both balanced and unbalanced power ratios) feedback into the gain section.}
\section{Results and discussions}
\subsection{Effects of feedback strength on RF linewidth and integrated timing jitter}
 {To investigate the effects of feedback on the RF linewidth and timing jitter, attenuation in the feedback loops was varied from -46 dB to -22 dB, after which the laser became unstable. At ~ -46 dB the RF linewidth was 73 kHz for single-loop, 69 kHz for unbalanced SDL and 75 kHz for balanced SDL feedback, with respective timing jitter 3, 2.9, 3.1 ps (10 kHz-100 MHz). At this weak feedback level, upon tuning of the optical delay, no deviation in the fundamental frequency occurs, and no major reduction in RF linewidth and timing jitter were seen relative to free-running. However, at higher feedback -29 dB, RF linewidth and timing jitter began to decrease: the former was 28.7 kHz for single-loop, 21.5 kHz for unbalanced SDL, and 29 kHz for balanced SDL; RMS timing jitter was 1.75 ps for single-loop, 1.6 ps for unbalanced SDL and 1.8 ps for balanced SDL feedback. Further increase in feedback ratio to -22 dB resulted in significant reduction in RF linewidth and timing jitter for single and SDL configuration subject to both balanced and unbalanced feedback ratios. Lowest achieved RF linewidth and timing jitter for single and SDL configurations as functions of feedback ratio at integer resonances are depicted in Figs. 2(a) and 2(b) respectively. From these data, we have identified the feedback ratio to be -22 dB for single-loop and SDL feedback, limited by self-pulsation above this level. For practical applications, the relatively flat characteristics of RF linewidth versus feedback ratio (-26 dB, -24 dB, -23 dB and -22 dB) are favorable. Furthermore, variation in RF linewidth and timing jitter in all feedback schemes studied follows similar trends when feedback approaches the optimal value, which agrees well with reported analytical results (square root dependence of the RF linewidth on integrated timing jitter) [28]. 
    \begin{figure}[ht!]
\centering\includegraphics[width=13cm]{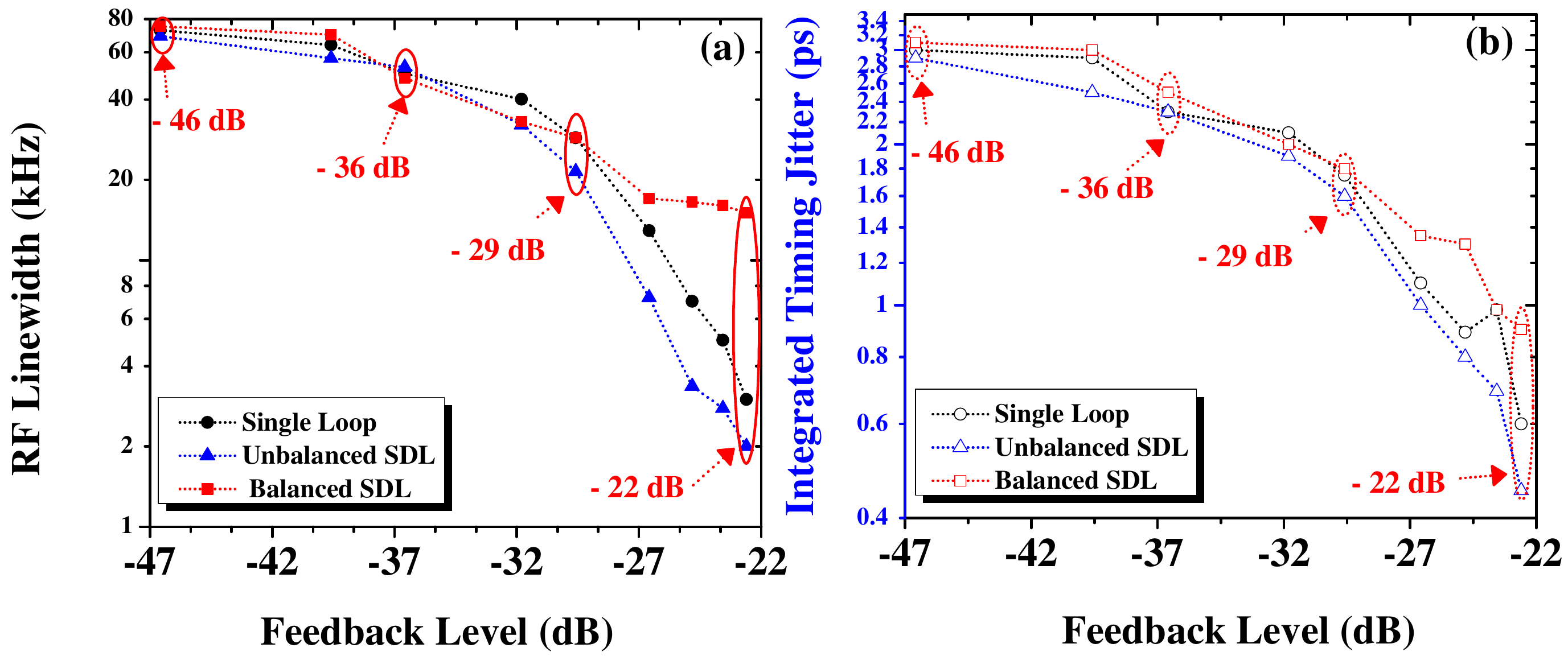}
\caption{(a) 3-dB RF linewidth and (b) integrated timing jitter under resonant condition subject to single-loop (black circles), unbalanced SDL (blue triangles) and balanced SDL  (red squares) feedback configurations as a function of external feedback ratio at a bias of 300 mA gain current }
\end{figure}

 Recently, for a quantum dot MLL operating at 5.1 GHz, minimum RF linewidth was obtained at relatively low feedback -36 dB [6]. On the other hand, a passively mode-locked QDash laser emitting at 1580 nm and operating at 17 GHz repetition rate saw marked reduction in RF linewidth at significantly stronger feedback -22 dB [7], in agreement with our studies. These differences are explicable by the likelihood that the anti-guiding (phase-amplitude coupling) factor, the primary measure of nonlinearity in these lasers, is lower in quantum dashes.}\par

\subsection{RF linewidth and integrated timing jitter versus delay for single-loop feedback}
{To study the effects of single-loop feedback on RF linewidth and timing jitter over a wide delay range (0-84 ps), loop-II was disconnected and maximum feedback to the gain section was set to -22 dB via a single 60 m fiber span, stable resonance being achieved by optimizing optical delay line ODL-I adjustable from 0-84 ps in steps of 1.67 ps. The resulting RF linewidth (black squares) and timing jitter (blue triangles) are shown in Fig. 3 versus delay. Clearly stabilization effectiveness depends strongly on feedback delay, most likely because detuning of optical delay from exact resonance changes synchronization conditions between pulses in the laser cavity and feedback loops [29]. The periodicity in RF linewidth versus delay tuning is 48 ps, in agreement with the fundamental mode-locked frequency (20.7 GHz) of our laser. Furthermore, this optimization of the single-loop delay reduced the RF linewidth and corresponding timing jitter considerably, as for other reported experiments [5,6] and theoretical predictions [30]. Effective stable mode-locking occurs when the external cavity optical length is close to an integer multiple of that of the laser cavity. When fully resonant, the RF linewidth decreased from 100 kHz free-running to 3 kHz, and integrated timing jitter from 3.9 ps to 0.6 ps (10 kHz-100 MHz). Measured RF spectra and phase noise traces at this feedback delay with single-loop feedback (blue line) and free-running (gray line) are shown in Figs. 4(a) and 4(b) respectively. Upon tuning of the loop delay by 6-54 ps, synchronization of the optical pulses between the laser cavity and external cavity did not occur, the RF spectra become highly deformed and non-resonant feedback was observed.}
  \begin{figure}[ht!]
\centering\includegraphics[width=10cm]{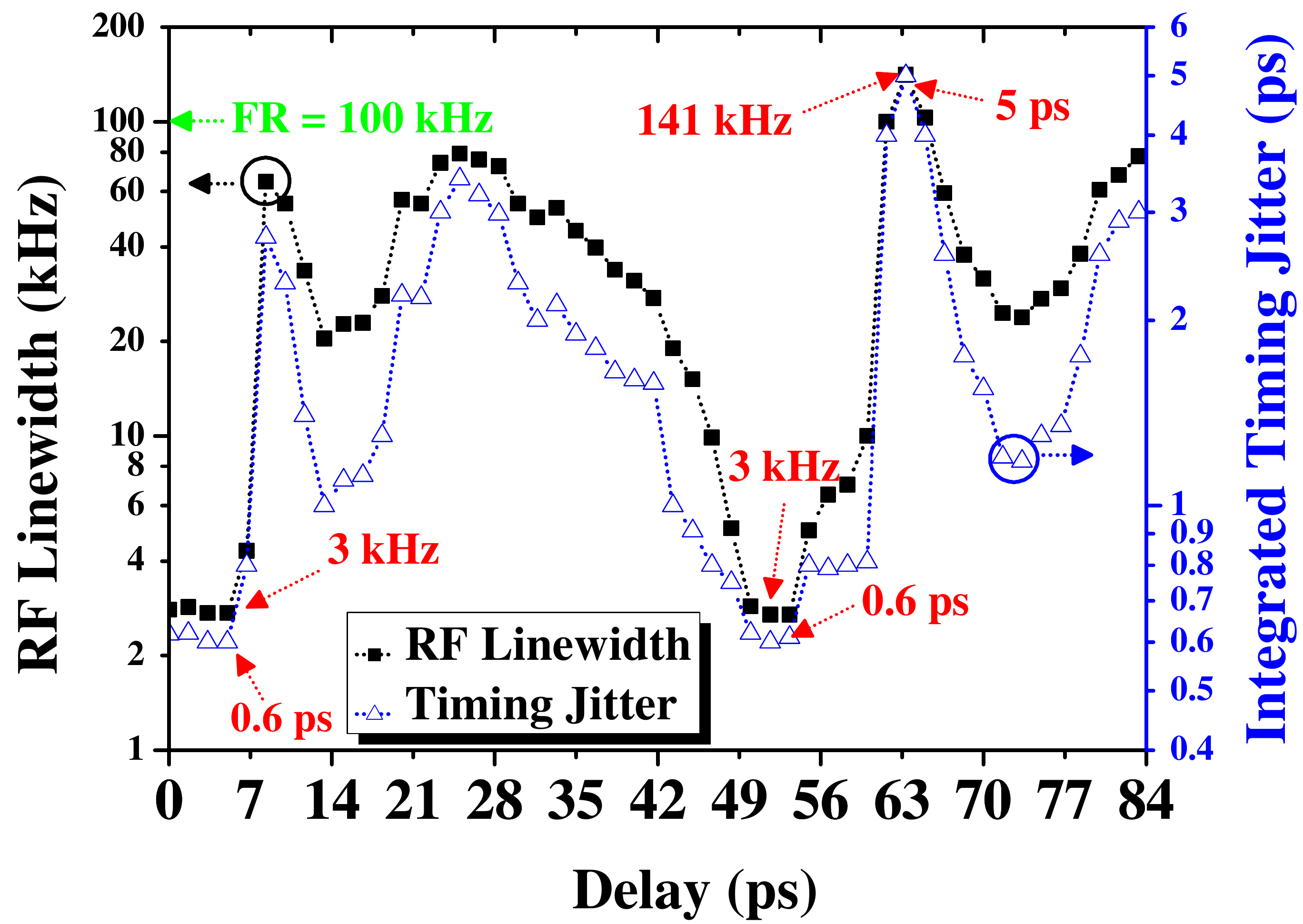}
\caption{RF linewidth (black squares) and integrated timing jitter (blue triangles) as functions
of delay, for single-loop optical feedback}
\end{figure}
 {Experimental results for single-loop feedback show that for practical use of QDash MLLs the most stable, delay-insensitive ranges are near 5 and 53 ps. Optimum stabilization using conventional single-loop feedback is very sensitive to phase adjustment, and limits the region of optimum performance to a narrow parameter space. For practical applications of MLLs, it is desirable to extend the range of resonant feedback condition over a much wider range of delay times, such that environment changes maintain stable pulse trains with narrow linewidth and low timing jitter.}
  \begin{figure}[ht!]\centering\includegraphics[width=13cm]{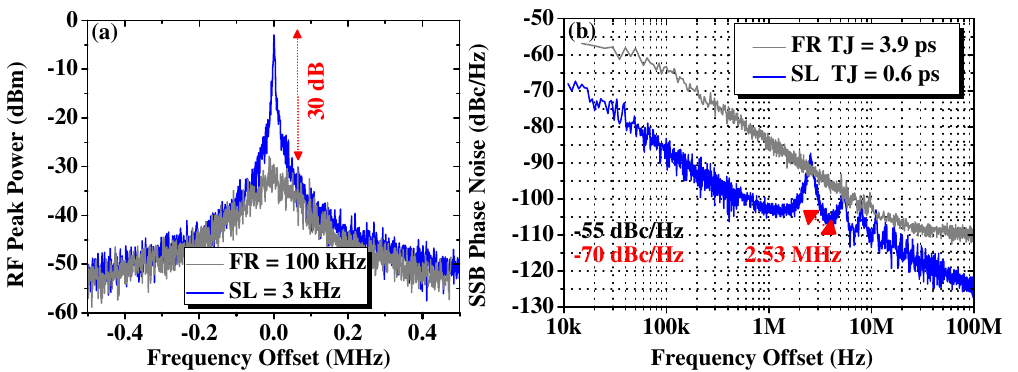}
\caption{Comparison of (a) RF linewidth and (b) phase noise traces of free-running (FR) laser (gray line) with single-loop (SL) feedback (blue line)}
\end{figure}

\subsection{RF linewidth and integrated timing jitter versus delay for balanced and unbalanced symmetric dual-loop (SDL) feedback arrangements}
 {For dual-loop experiments, the optical feedback was split into two fiber cavities whose lengths were calibrated by measurement of RF spectra with each loop unblocked separately. For the first set, cavity spacing was 2.53 MHz consistent with 80 m nominal length of both equal loops. RF spectral measurements with this arrangement (using frequency span 10 MHz with resolution bandwidth 10 kHz and video bandwidth 1 kHz) are shown in Fig. 5(a). 
}\par
     \begin{figure}[ht!]
\centering\includegraphics[width=13cm]{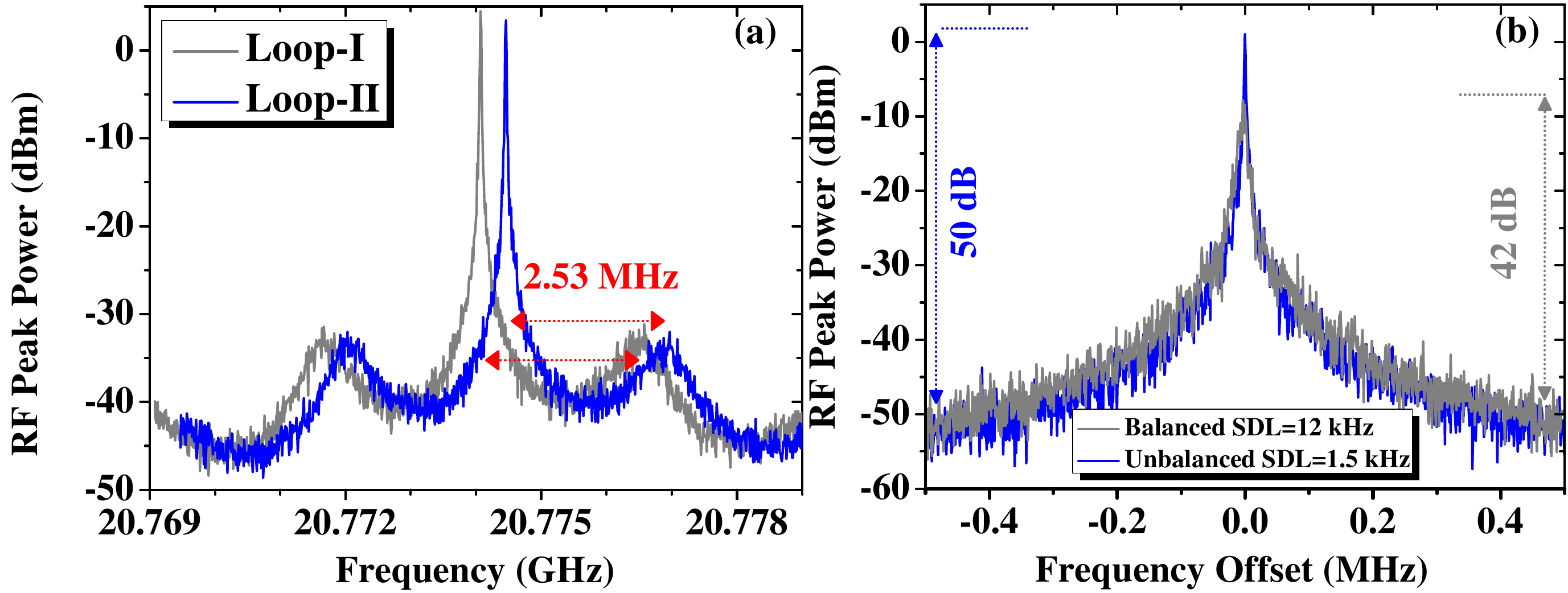}
\caption{(a) Separately measured RF spectra of single-loop feedback from loop-I (gray line) and loop-II (blue line) (b) Comparison of RF spectra under resonant condition for balanced (gray line) and unbalanced (blue line) SDL feedback}
\end{figure}
{To study the effects of SDL feedback on laser stability, fine adjustment of optical attenuator (Att-I) and polarization controller (PC-I) was made and equal feedback power (-22 dB) coupled to both loops. Optical delay line ODL-I was set to full resonance (integer number of times the laser cavity delay) and ODL-II tuned over its full range. RF linewidth (black squares) and timing jitter (black squares) are presented as functions of delay in Figs. 6(a) and 6(b). We see that SDL feedback yields results comparable to those using a single-loop, and is similarly sensitive to delay. Only with both loops resonant does stability improve, giving RF linewidth 12 kHz and timing jitter 0.85 ps, versus 3 kHz and 0.6 ps for optimized single-loop feedback, refer to Fig. 5(b). Recently, in a separate series of experiments, we achieved 0.97 kHz linewidth (instrument limited) and timing jitter 0.45 ps with both cavities resonant [17], confirming that balanced SDL produces effective stabilization, but only at a specific delay with tolerance $\sim$ 1 ps, a very stringent requirement in practice.}\par
\begin{figure}[ht!]
\centering\includegraphics[width=13cm]{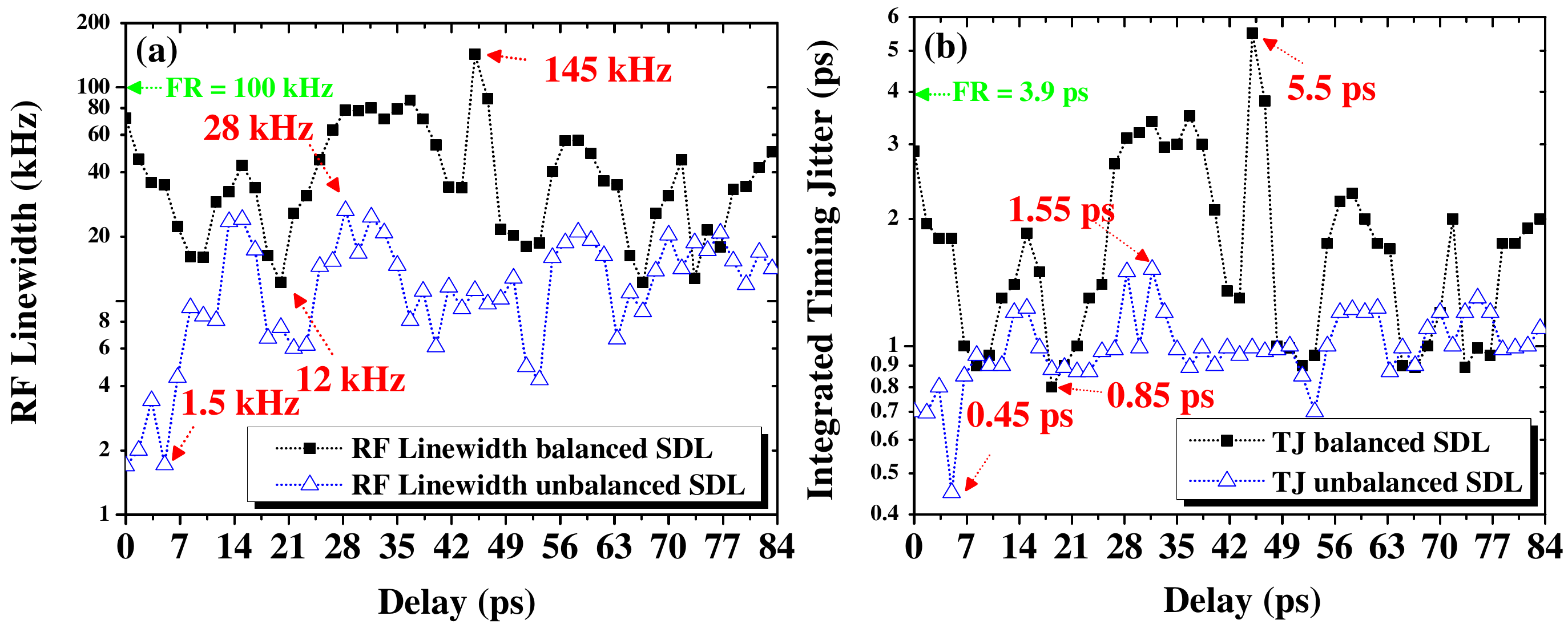}
\caption{(a) RF linewidth and (b) Integrated timing jitter as a function of full delay phase subject to balanced (black squares) and unbalanced (blue triangles) SDL feedback configuration}
\end{figure}  
 {Next, we explored unbalanced SDL feedback, in which the power split between the two loops was varied. In these experiments the inner feedback cavity (loop-I) was fully resonant and the outer feedback cavity fine-tuned around the resonance. Measured RF spectra at different loop power splits are shown in Fig. 7, with corresponding RF linewidths in Table 1. Minimum RF linewidth occurred when both external cavities were fully resonant, as expected. Values of 1.6 and 1.5 kHz were achieved when resonant loop-I had feedback -20.6 and -20 dB, and fine-tuned loop-II had -24.3 and -26 dB, respectively. This combination of feedback ratios was particularly effective and was investigated further.}\par
 \begin{figure}[ht!]
\centering\includegraphics[width=13cm]{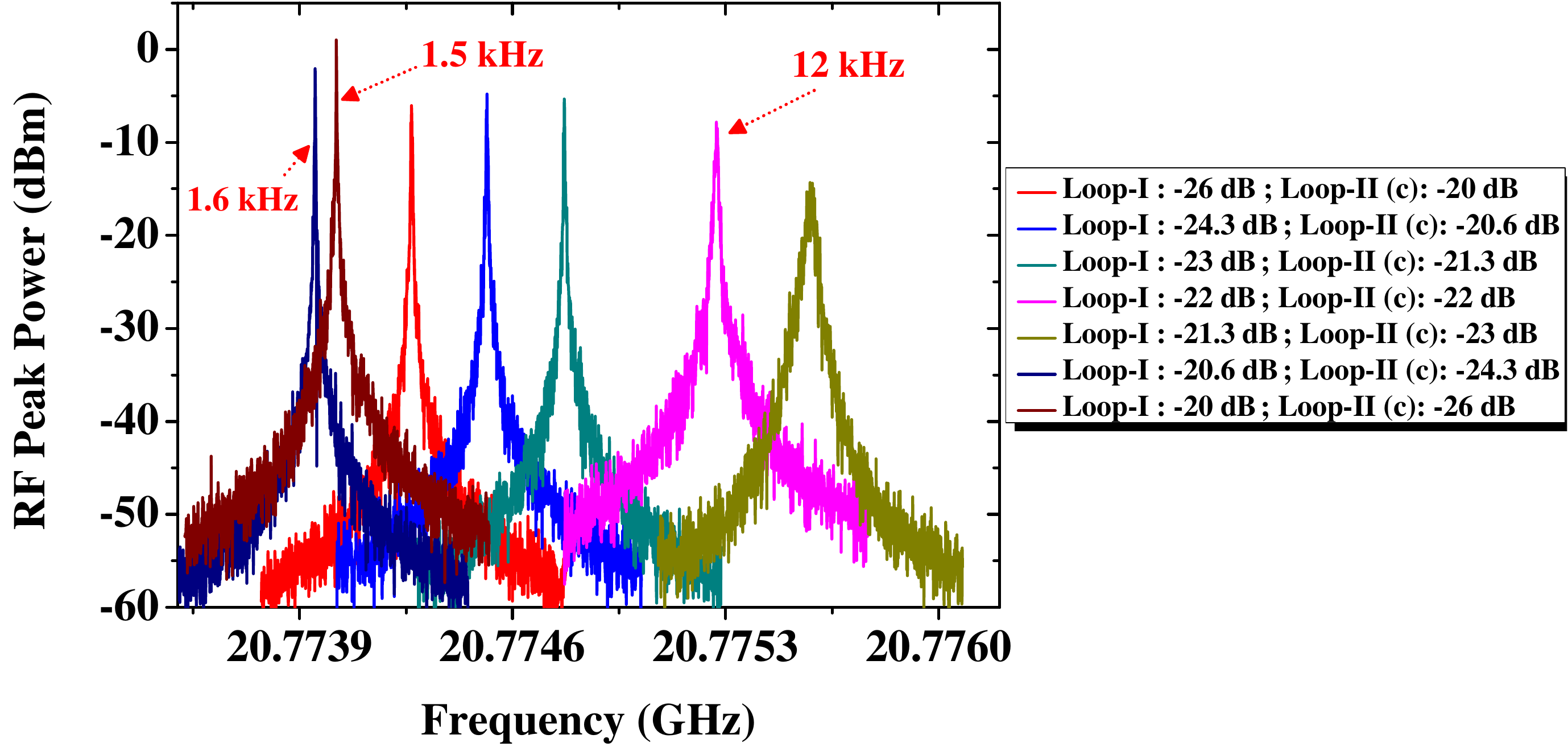}
\caption{Measured RF spectra at different feedback ratios for symmetric dual-loops. (Frequency span 1 MHz, resolution bandwidth 1 kHz, video bandwidth 100 Hz}
\end{figure}  
\begin{table}[ht!]
\centering
\caption{Calculated RF linewidth as a function of power split ratio (in dB) through two external feedback loops using SDL feedback configuration}
\begin{tabular}{|l|l|l|l|}
\hline
\textbf{Loop-I} & \textbf{Loop-II (c)} &  \textbf{Feedback into Gain} & \textbf{RF-Linewidth}\\ \hline
-26 dB & -20 dB & -22 dB&4.1 kHz \\ \hline
-24.3 dB & -20.6 dB& -22 dB&3.4 kHz \\ \hline
-23 dB & -21.3 dB& -22 dB&2.1 kHz \\ \hline
-22 dB & -22 dB &-22 dB& 12 kHz \\ \hline
-21.3 dB& -23 dB & -22 dB&30 kHz \\ \hline
 -20.6 dB& -24.3 dB& -22 dB&1.6 kHz \\ \hline
-20 dB& -26 dB& -22 dB&1.5 kHz \\ \hline
\end{tabular}
\end{table}
{The next experiments concerned the effects of unbalanced SDL feedback on timing stability of the laser: feedback strength in loop-I and loop-II were -20 and -26 dB, a ratio of 4:1 resulting in overall feedback -22 dB to the gain section. Delay in loop-I was then fine-tuned to full resonance, and loop-II tuned over its entire available delay range 0-84 ps. This yielded much more stable dynamics: narrow RF spectra and reduced timing jitter were maintained over the full delay range, unlike single-loop and balanced SDL feedback. This dual-loop scheme with 4:1 power ratio between loops was most successful, reducing RF linewidth by up to two orders of magnitude (70x) compared to free-running, 2-5x over single-loop and 5-8x relative to balanced SDL feedback. Measured RF linewidths (blue triangles) and timing jitter (blue triangles) for this unbalanced SDL scheme are given in Figs. 6(a) and 6(b). Furthermore, with this feedback configuration, measured RF linewidth and integrated timing jitter ranged from as high as 28 kHz and 1.5 ps, to as low as 1.5 kHz and 0.45 ps (free running values are 100 kHz and 3.9 ps). Again, most effective and robust linewidth narrowing and lowest timing jitter occurred when both external cavities were fully resonant. The RF spectrum under double resonance is shown in Fig. 5(b) (blue line). RF spectra (with 10 MHz frequency span, 10 kHz resolution bandwidth and 1 kHz video bandwidth) and phase noise for unbalanced SDL feedback versus frequency offset are given in Figs. 8(a) and 8(b). Side-modes were spaced by 2.53 MHz, corresponding to the feedback loop length used (80 m). Most recently, we have used asymmetric dual-loop feedbacks with optimized inner loop delay to suppress spurious tones and timing jitter in self-mode-locked lasers [31]. Furthermore, when free-running the peak power of RF noise spectra is -20 dB. For single-loop and unbalanced SDL feedback the noise peak is 30 dB higher (see Fig. 4(a) and Fig. 5(b)) due to the reduced RF linewidth and also lower threshold current with feedback, increasing the optical power emitted at fixed bias. } \par
     \begin{figure}[ht!]
\centering\includegraphics[width=13cm]{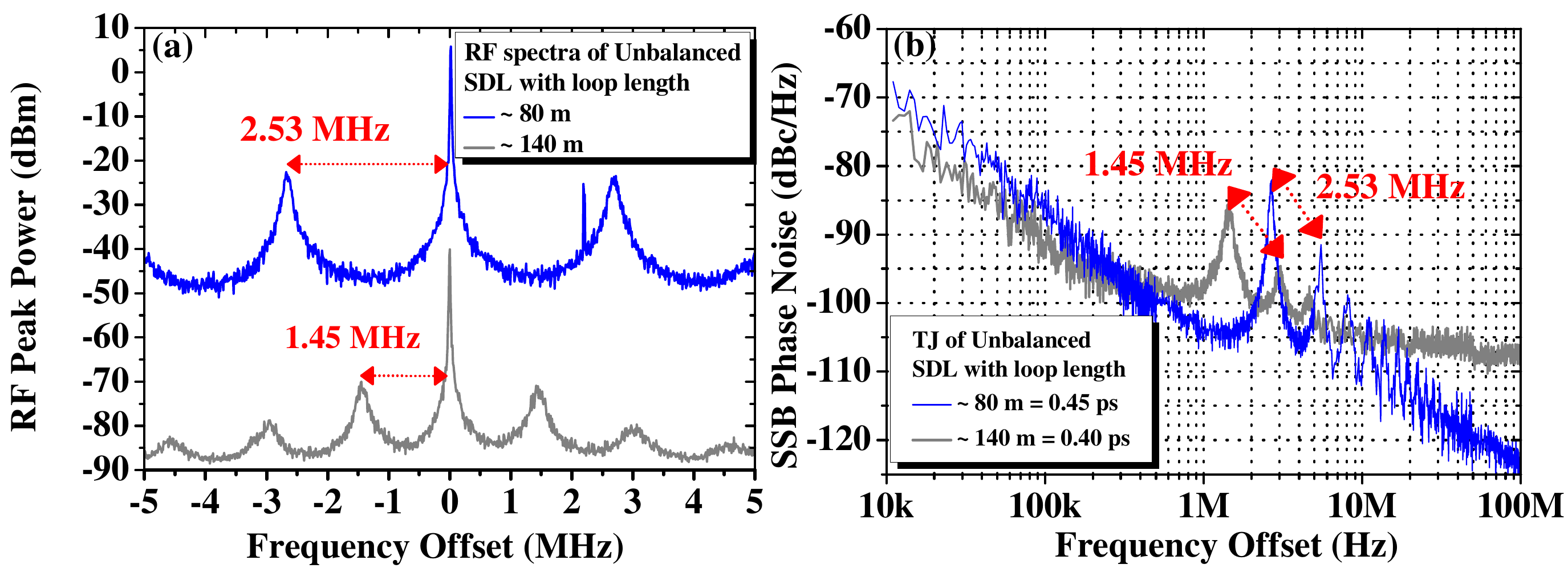}
\caption{Comparison of measured (a) RF spectra and (b) phase noise trace between loop length 80 m (blue line) and 140 m (gray line) using unbalanced SDL feedback configuration}
\end{figure}
{Measured RF linewidths versus delay for a single-loop, with -20 dB feedback through loop-I (blue triangles), are shown in Fig. 9(a) for comparison. At stable resonance, single-loop feedback at -20 and -26 dB narrows the linewidth to 8 kHz and 68 kHz, respectively. When dual-loops were unbalanced, measured RF linewidth as a function of delay was as in Fig. 9(b), showing that unbalanced dual-loops are more effective in stabilizing the linewidth. Here optimization of ODL-II yields better linewidth stabilization (blue triangles) than optimization of ODL-I (black squares), see Fig. 9(b). For SDL feedback, fine tuning of ODL-I yields narrow RF linewidth at an integer resonance, but linewidth broadens significantly when delay is tuned away from this point. Our results show that the most effective algorithm for stable linewidth reduction over a broad range of phase delay is to set the stronger cavity to an integer resonance then fine-tune the weaker cavity. Optimizing loop-II in SDL feedback (blue triangles in Fig. 9(b)), changes the linewidth similarly to single-loop feedback (black squares in Fig. 9(a)) but almost 1-2 orders of magnitude (6-64x) narrower.}\par
     \begin{figure}[ht!]
\centering\includegraphics[width=13cm]{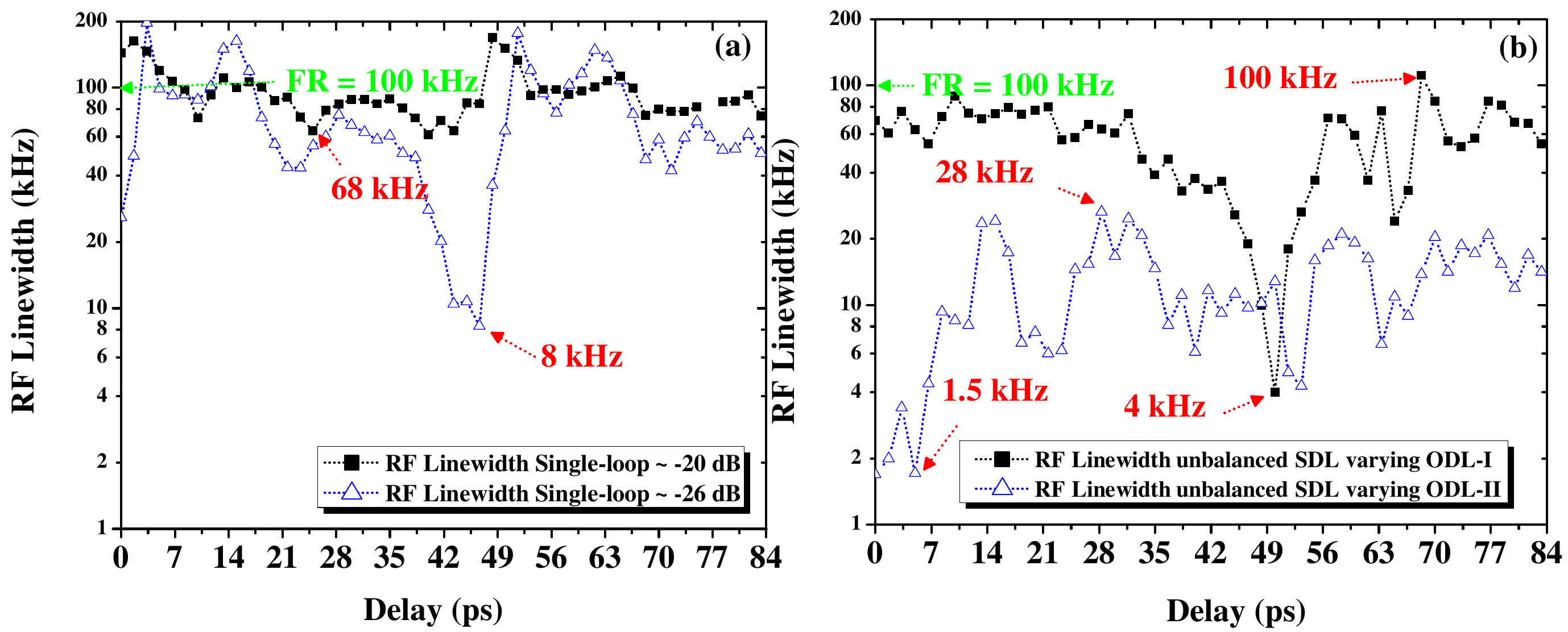}
\caption{(a) RF linewidth as a function of delay using single-loop feedback with strength -20 (blue triangles) and -26 dB (black squares) (b) RF linewidth for unbalanced SDL feedback configuration with optimization of ODL-I (black squares) and ODL-II (blue triangles)}
\end{figure}
{Effects of longer delay times on RF linewidth were also investigated subject to both balanced and unbalanced SDL feedback. For this purpose, the 60 m fiber loop was replaced with 120 m fiber. Measured RF linewidth versus phase tuning is shown in Fig. 10 for SDL with balanced (black squares) and unbalanced (blue triangles) feedback ratios. Unbalanced SDL feedback again reduced the RF linewidth by 10-100x across a broader range of delay than the free-running laser. However, balanced SDL was less sensitive to delay, though the linewidth was 8-16x broader than unbalanced SDL (Fig. 10). Narrowest linewidth obtained was 1 kHz under full resonance (25 ps delay in Fig. 10) which is the limit of resolution of our spectrum analyzer. Timing jitter was also minimized at 0.4 ps. The RF spectrum under these conditions is shown in Fig. 8(a) (gray line). We observed 1.45 MHz external cavity mode spacing as expected for a 140 m loop. Measured phase noise versus frequency offset from the fundamental mode-locked frequency is shown in Fig. 8(b) (gray line). RF linewidth and timing jitter were lower over a wider delay range with the longer cavity, due to its higher quality factor (Q).}\par
       \begin{figure}[ht!]
\centering\includegraphics[width=10cm]{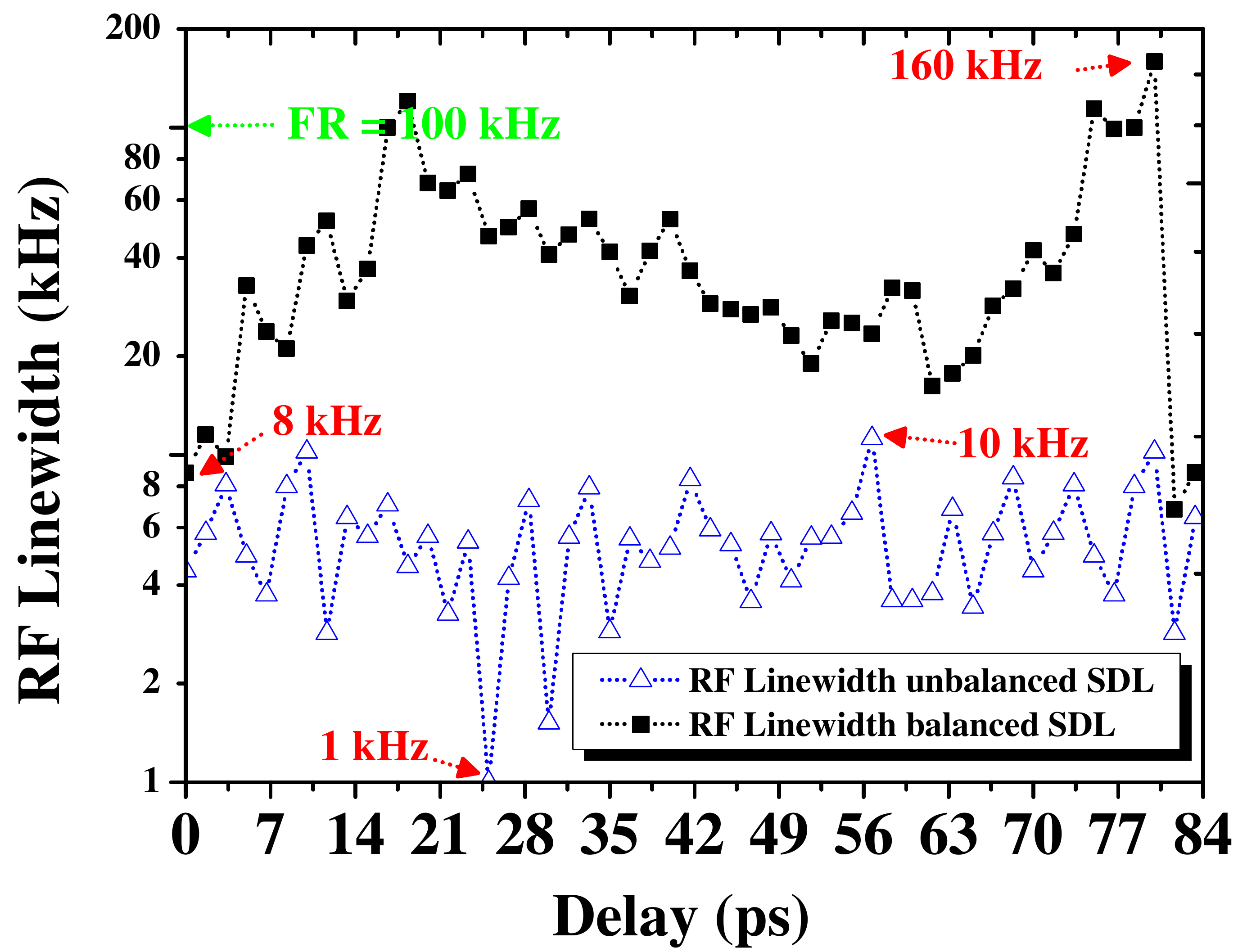}
\caption{RF linewidth subject to unbalanced (blue triangles) and balanced (black squares) SDL feedback configuration as a function of maximum available delay phase tuning (0-84 ps)}
\end{figure}
{In summary, our experiments demonstrate that unbalanced SDL is more effective than balanced SDL and single-loop feedback in stabilizing the laser over a wider delay range. Recent experiments have shown asymmetric dual-loop feedback improves timing jitter of a QDash MLL and  suppresses unwanted spurious side-bands. With this resonant arrangement, sub-kHz RF linewidth, sub-picosecond integrated timing jitter and 30 dB side-mode suppression were simultaneously achieved. Recently, it was theoretically predicted that dual-loop optoelectronic oscillators could be optimized by controlling the phase delay and power split ratio [32]. Our unbalanced (asymmetric power split) SDL configuration produces narrower linewidth, lower timing jitter over a delay range limited to 0-84 ps by the variable delay lines available, with only one loop requiring fine-tuning. This promises to be a robust and effective means to stabilize mode-locked QDash lasers emitting at $\sim$ 1550 nm.}\par
    \section{Conclusion}
{We investigated the effectiveness of single-loop and SDL optical feedback as means of robust stabilization of self-mode-locked QDash lasers operating at 21 GHz pulse repetition rate and emitting at 1550 nm wavelength. Mode-locking occurred without reverse bias applied to the absorber section, and robust stabilization was achieved with predictable delay difference between the two external feedback cavities, which simplifies product design and packaging. We demonstrated that unbalanced SDL feedback provides best stability, maintaining stable RF spectra with narrow linewidth and low timing jitter over a range of delay detuning ~80 ps, which means it would be insensitive to temperature, vibration and other common environmental variations. Unbalanced SDL is significantly better than conventional single-loop feedback and balanced SDL feedback, producing up to two orders of magnitude reduction in RF linewidth to 1 kHz (instrument limited) and RMS timing jitter 0.4 ps, compared to free-running. Longer 140 m fiber loops are more effective than shorter 80 m loops. For SDL feedback, we have studied the effects of varying the power split between the loops. The proposed scheme is effective in overcoming the primary drawback of mode-locked diode lasers, their lack of dynamical stability and robustness, in practical applications such as frequency comb generation, optical sampling, signal timing and regeneration, metrology, lidar and many others.}
\section*{Acknowledgments}
The authors acknowledge financial support from Science Foundation Ireland (grant 12/IP/1658) and the European Office of Aerospace Research and Development (grant FA9550-14-1-0204). 
\end{document}